\def\lapprox{\mathrel{\mathop
  {\hbox{\lower0.5ex\hbox{$\sim$}\kern-0.8em\lower-0.7ex\hbox{$<$}}}}}
\def\gapprox{\mathrel{\mathop
  {\hbox{\lower0.5ex\hbox{$\sim$}\kern-0.8em\lower-0.7ex\hbox{$>$}}}}}

\documentstyle[iopconf1,epsf]{article}

\begin{document}
\title{
\vbox to 0pt{{\ }\par\vskip-2.5cm \noindent
\normalsize Contribution to the Proceedings of 
{\it Beyond the Desert}, Ringberg Castle, Tegernsee, Germany
June 8--14, 1997.\vfil}
\vskip -24pt
Astrophysical axion bounds: An update}

\author{Georg G Raffelt\footnote{E-mail: raffelt@mppmu.mpg.de}}

\affil{Max-Planck-Institut f\"ur Physik (Werner-Heisenberg-Institut),
F\"ohringer Ring 6, 80805 M\"unchen, Germany}

\beginabstract 

The observed properties of stars and especially the neutrino signal
of the supernova 1987A provide an upper limit to the axion mass, while
the age and expansion rate of the universe provide a lower limit.
There remains a ``window of opportunity'' $10^{-5}\,{\rm
eV}\lapprox m_a\lapprox 10^{-2}\,{\rm eV}$, with large uncertainties
on either side, where axions could still exist and where they would
provide a significant fraction or all of the cosmic dark matter.
The current status of this axion window is reviewed.

\endabstract

\section{Introduction}

The interest in axions as a possible dark matter candidate has
recently soared thanks to the heroic experimental progress which has
led to full-scale searches for galactic axions in Livermore
\cite{Livermore} and Kyoto~\cite{Kyoto} which may well turn up
dark matter axions within the next few years (Fig.~1).  
There is also a noteworthy axion search program in
Novosibirsk~\cite{Novosibirsk}.

For such efforts to make sense one needs to understand the ``window
of opportunity'' where axions are not excluded by astrophysical and
cosmological arguments. It is well known that the requirement that
stars not lose too much energy by axions leads to a lower limit on the
Peccei-Quinn scale $f_a$ which can be translated into an upper limit
on the axion mass by virtue of the relationship $m_a=0.62\,{\rm
eV}\,(10^7\,{\rm GeV}/f_a)$.  It is also well known that the
non-thermal production in the early universe leads to an upper bound
on $f_a$ (lower bound on $m_a$) lest axions ``overclose'' the
universe. These topics are well-covered in a number of reviews
\cite{reviews} and books~\cite{KolbTurner,RaffeltBook}. Some recent
refinements, however, justify the present update.

\begin{figure}[t]
\hbox to \hsize{\hfil
\epsfxsize=7.9cm
\epsfbox{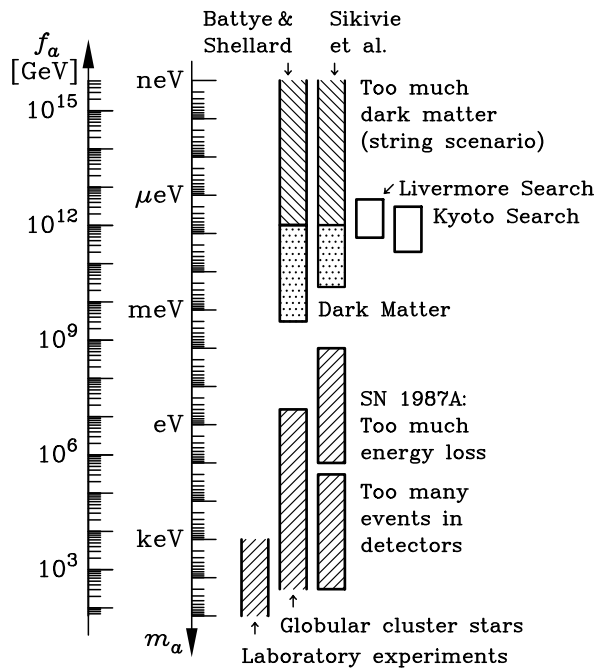}\hfil}
\caption{Summary of astrophysical and cosmological axion bounds.
The globular cluster limit assumes an axion-photon coupling
corresponding to $E/N=8/3$ as in GUT models.}
\end{figure}

\section{Stellar Limits}

\subsection{Globular Cluster Stars}

In analogy to neutral pions, axions generically interact with photons
according to ${\cal L}_{a\gamma}=g_{a\gamma} {\bf E}\cdot{\bf B}\,a$
with $g_{a\gamma}=(\alpha/2\pi f_a)\,(E/N-1.92\pm0.08)$.  The
parameter $E/N$ is a model-dependent fraction of small integers.  This
coupling allows for the axion decay $a\to2\gamma$ as well as for the
Primakoff conversion $a\leftrightarrow\gamma$ in the presence of
external electromagnetic fields. Because charged particles and photons
are abundant in the interior of stars they are powerful axion sources.

A novel energy-loss mechanism would accelerate the consumption of
nuclear fuel in stars and thus shorten their lifetimes.  A case where
axion emission would be efficient and the stellar lifetime is well
established are low-mass helium-burning stars, so-called
horizontal-branch (HB) stars.  Low-mass red giants have a degenerate
helium core ($\rho\approx10^6\,\rm g\,cm^{-3}$, $T\approx10^8\,\rm K$)
so that axion emission is strongly suppressed relative to the cores of
HB stars ($\rho\approx10^4\,\rm g\,cm^{-3}$, $T\approx10^8\,\rm K$)
whence the number ratio of these stars in globular clusters is a
sensitive measure for the operation of axionic energy losses. The
observed number ratios agree with standard theoretical expectations to
within a few tens of percent implying a limit~\cite{RaffeltBook}
$g_{a\gamma}\lapprox0.6{\times}10^{-10}\,{\rm GeV}^{-1}$.  In GUT
axion models where $E/N=8/3$ this yields $m_a\lapprox 0.4\,{\rm eV}$
(Fig.~1).  Our often-quoted ``red giant limit'' is less restrictive
because it was based on the statistically less significant number of
``clump giants'' in open clusters~\cite{Dearborn}.

\subsection{Supernova 1987A}

Being a QCD phenomenon, axions $a$ generically also couple to nucleons
$N$ by $(C_N/2f_a) \overline\psi_N\gamma_\mu\gamma_5\psi_N 
\partial^\mu a$ where the $C_N$ for protons and neutrons
are model-dependent numerical coefficients.  The most significant
limit on the axion-nucleon coupling arises from the cooling speed of
nascent neutron stars as established by the duration of the neutrino
signal from the supernova (SN) 1987A.

Apart from the well-known overall uncertainty of this argument caused
by the statistical weakness of only 19 observed neutrinos there is a
problem related to the axion emission rate from the hot and dense
nuclear medium.  In a naive perturbative picture this rate is computed
as a bremsstrahlung process $NN\to NN a$ with the nucleons interacting
by a spin-dependent force. The resulting nucleon spin fluctuations act
as a source for the emission of axions. Like any other bremsstrahlung
process, this rate scales with the density as $\rho^2$. However, if
one studies axion emission from the more general perspective of linear
response theory it turns out that the $\rho^2$ scaling is not
maintained to arbitrary densities. Rather, the axion emission rate
does not exceed a certain limit which can be estimated from sum rules
of the spin-density dynamical structure functions~\cite{JKRS,Sigl}.
While a detailed calculation of these structure functions is not
available, simple estimates indicate that in a SN core one is well in
the saturation regime of the bremsstrahlung process. The previous
naive emission rate is thus reduced by about an order of magnitude.

Moreover, a previous often-quoted limit \cite{BTR} of $m_a\lapprox
1\,{\rm meV}$ was based on the generic coupling constants $C_N=0.5$
for both protons and neutrons.  In realistic axion models these
couplings can be much smaller. For example, in the popular KSVZ model
which is representative for the class of hadronic axion models the
interaction with neutrons nearly vanishes. Altogether the SN~1987A
limit is reduced to roughly $m_a\lapprox 10\,{\rm meV}$ \cite{JKRS}.
Inevitably this limit involves large uncertainties which are difficult
to quantify.

Most recently the SN 1987A axion limit was reexamined in a series of
numerical cooling calculations where the saturation effect was
included and the coupling constants to neutrons and protons were
chosen appropriately for specific models \cite{Keil}.  For KSVZ axions
the limit was found to be about $8\,{\rm meV}$, in perfect agreement
with the simple estimate of Ref.~\cite{JKRS}. For DFSZ axions the
limit varies between about 4 and $12\,{\rm meV}$, depending on the
assumed value of the angle $\beta$ which measures the ratio of two
Higgs vacuum expectation values. Because this angle is not known one
can only use the least restrictive number as a conservative limit. In
view of the large overall uncertainties it is good enough to remember
$m_a\lapprox 10\,\rm meV$ as the current SN 1987A limit for both KSVZ
and DFSZ axions.

If axions interact too strongly they are trapped and contribute to the
transfer of energy rather than to a direct cooling of the inner SN
core.  Therefore, axions with $m_a\gapprox 10\,{\rm eV}$ cannot be
excluded on the basis of the duration of the SN~1987A neutrino 
signal~\cite{Burrows}.

However, axions with masses larger than this, i.e., with stronger
interactions, could actually cause a significant contribution to the
signal measured in the IMB and Kamiokande~II water Cherenkov detectors
by the absorption on $^{16}\rm O$ and the subsequent emission of
$\gamma$ rays. To avoid too many events one can exclude the
range $20\,{\rm eV}\lapprox m_a\lapprox 20\,{\rm keV}$
\cite{Engel}.

\subsection{White Dwarf Cooling}

In certain models axions couple to electrons by 
$(C_e/2f_a)\overline\psi_e\gamma_\mu\gamma_5\psi_e
\partial^\mu a$ with $C_e$ a model-dependent factor of order unity.
For most purposes this derivative coupling is equivalent to the
pseudoscalar structure 
$-ig_{ae}\overline\psi_e\gamma_5\psi_e a$ with the Yukawa
coupling $g_{ae}=C_e m_e/f_a$ which one may parametrize
by $\alpha_{26}\equiv (g_{ae}^2/4\pi)/10^{-26}$.

It was suggested that axion emission with $\alpha_{26}\approx0.45$
might dominate the cooling of white dwarfs such as the ZZ~Ceti star
G117--B15A for which the cooling speed has been established by a
direct measurement of the decrease of its pulsation period
\cite{Isern}. Because of this suggestion a new bound on $\alpha_{26}$
was derived by a method similar to the above number counts in globular
clusters \cite{RaffeltWeiss}. The resulting limit
$\alpha_{26}\lapprox0.5$ is the best direct bound on the
axion-electron coupling, but it does not quite exclude the possibility
that axions could play a certain role in white dwarf cooling.

The most popular example where axions couple to electrons is the DFSZ
model where $C_e={1\over3}\cos^2\beta$ with $\beta$ an arbitrary
angle.  In this case one may use the SN 1987A limits on the
axion-nucleon coupling to derive an indirect $\beta$-dependent limit
on the axion-electron coupling. In this way I infer from
Ref.~\cite{Keil} that the largest axion-electron coupling allowed by
SN~1987A corresponds to $\alpha_{26}\approx 0.08$. For typical
parameters of old white dwarfs the axion luminosity is
$0.7\,\alpha_{26}$ times their photon
luminosity~\cite{RaffeltBook}. Therefore, DFSZ axions do not seem to
be able to play a significant role in the cooling of old white
dwarfs. (A novel cooling mechanism may be required if the microlensing
events of the MACHO collaboration are to be interpreted as halo white
dwarfs~\cite{Freese}.) However, axion cooling may be important in
strongly magnetized white dwarfs where the cyclotron process provides
for an additional emission channel~\cite{Kachelriess}.

\section{Mass of Dark Matter Axions}

If axions interacted sufficiently strongly ($f_a\lapprox10^8\,{\rm
GeV}$) they would have come into thermal equilibrium before the QCD
phase transition, leading to a background sea of invisible axions in
analogy to the one expected for neutrinos \cite{TurnerI}. This
parameter range is excluded by the astrophysical arguments summarized
in Fig.~1 which imply that axions interact so weakly that they have
never come into thermal equilibrium.  Still, the well-known
misalignment mechanism will excite coherent oscillations of the axion
field~\cite{Misalignment}. In units of the cosmic critical density one
finds for the axionic mass density
\begin{equation}
\Omega_a h^2\approx1.9\times 4^{\pm1} 
(\mu{\rm eV}/m_a)^{1.175}\,
\Theta_i^2\,F(\Theta_i)
\end{equation} 
where $h$ is the present-day Hubble expansion parameter in units of
$100\,\rm km\,s^{-1}\,Mpc^{-1}$. The stated range reflects recognized
uncertainties of the cosmic conditions at the QCD phase transition and
uncertainties in the calculation of the temperature-dependent axion
mass.  The cosmic axion density depends on the initial misalignment
angle $\Theta_i$ which could lie anywhere between 0 and $\pi$. The
function $F(\Theta)$ encapsules anharmonic corrections to the axion
potential; for an analytic determination see Ref.~\cite{Weiler}.

If $\Theta_i$ is of order unity, axions with $m_a={\cal O}(1\,\mu{\rm
eV})$ provide roughly the cosmic closure density. The equivalent
Peccei-Quinn scale $f_a={\cal O}(10^{12}\,{\rm GeV})$ is far below the
GUT scale so that one may speculate that cosmic inflation, if it
occurred at all, did not occur after the PQ phase transition.  If it
did not occur at all, or if it did occur before the PQ transition with
$T_{\rm reheat}>f_a$, the axion field will start with a different
$\Theta_i$ in each region which is causally connected at $T\approx
f_a$. Then one has to average over all regions to obtain the
present-day axion density.

More importantly, because axions are the Nambu-Goldstone mode of a
complex Higgs field after the spontaneous breaking of a global U(1)
symmetry, cosmic axion strings will form by the Kibble
mechanism~\cite{Davis}.  The motion of these global strings is damped
primarily by the emission of axions rather than gravitational
waves. At the QCD phase transition the U(1) symmetry is explicitly
broken (axions acquire a mass) causing domain walls bounded by strings
to form which get sliced up by the interaction with strings. The whole
string and domain-wall system will quickly decay into axions. This
complicated sequence of events leads to the production of the dominant
contribution of cosmic axions. Most of them are produced near the QCD
transition at $T\approx\Lambda_{\rm QCD}\approx 200\,{\rm MeV}$. After
they acquire a mass they are nonrelativistic or mildly relativistic so
that they are quickly redshifted to nonrelativistic velocities. Thus,
the string and domain-wall produced axions form a cold dark
matter component.

In their recent treatment of axion radiation from global strings,
Battye and Shellard~\cite{BattyeShellard} found that the dominant
source of axion radiation are string loops rather than long strings,
contrary to what had been assumed in the previous works by
Davis~\cite{Davis} and Davis and Shellard~\cite{DavisShellard}.  At a
given cosmic time $t$ the average loop creation size is parametrized
as $\langle\ell\rangle=\alpha t$ while the radiation power from loops
is $P=\kappa\mu$ with $\mu$ the renormalized string tension. The 
loop contribution to the cosmic axion density is~\cite{BattyeShellard}
\begin{equation}
\Omega_a h^2\approx 88\times 4^{\pm1}
\left[(1+\alpha/\kappa)^{3/2}-1\right]\,(\mu{\rm eV}/m_a)^{1.175},
\end{equation}
where the overall uncertainty has the same source as before.  The
exact values of the parameters $\alpha$ and $\kappa$ are not known;
Battye and Shellard expect $0.1<\alpha/\kappa<1.0$. The expression in
square brackets is then between 0.15 and 1.83.

The proper treatment of axion radiation by global strings has been
controversial.  Sikivie and his
collaborators~\cite{SikivieHagmann} have consistently argued that the
motion of global strings was overdamped, leading to an axion spectrum
emitted from strings or loops with a flat frequency spectrum. In
Battye and Shellard's treatment, wavelengths corresponding to the loop
size are strongly peaked; the motion is not overdamped. In Sikivie
et~al.'s picture much more of the string-radiated energy goes into
kinetic axion energy which is redshifted so that ultimately there are
fewer axions. The cosmic axion density is then of
order the misalignment contribution.

If axions are the dark matter of the universe one may estimate
$0.08<\Omega_a h^2<0.40$. I have assumed that the universe is older
than $10\,\rm Gyr$, that the total matter density is dominated by
axions with $0.3<\Omega_{\rm matter}<1$, and that
$0.5<h<1.0$. Including all of the previously discussed uncertainties I
arrive at a plausible mass range for dark-matter axions of
\begin{equation}
m_a=\cases{\hbox{6--250 $\mu$eV}& Sikivie et al.,\cr
\hbox{6--2000 $\mu$eV}& Battye and Shellard,\cr}
\end{equation}
as indicated in Fig.~1. Even though Battye and Shellard tend to favor
larger axion masses for the dark matter, the treatments of both groups
imply the same lower end for the plausible
range. Evidently, the Livermore and Kyoto search experiments cover a
large fraction of the predicted range, even though it would be
desirable to extend the search to even larger masses.

\section{Phase Space Distribution}

The galactic phase-space distribution of axions may well exhibit novel
features which are of relevance for the search experiments.  Different
initial misalignment angles in different causally connected regions
lead to density fluctuations which are nonlinear from the start.  This
leads to the formation of ``axion mini clusters'' which may partially
survive galaxy formation and thus can be found in the Milky Way
today~\cite{Rees}.  For a suitably large initial density contrast
these clusters can condense into axionic boson stars by virtue of
higher order axion-axion couplings and may be detectable by the
femtolensing effect~\cite{Kolb}. The direct search experiments need
sufficient sensitivity to pick up the diffuse component of the
galactic axions which are not locked up in mini clusters.

Another interesting possibility is that axions may have maintained
some of their initial phase-space distribution, i.e.\ that they are
not fully virialized in the galaxy. In this case the axion velocity
distribution would exhibit very narrow peaks which could enhance the
sensitivity of the direct search experiments and in any case may be
detectable in the laboratory~\cite{Ipser}.

\section{Summary}

After some refinements and corrections, the astrophysical and
cosmological axion limits seem to have stabilized to what is shown in
Fig.~1. There remains a ``window of opportunity'' $10^{-5}\,{\rm
eV}\lapprox m_a\lapprox 10^{-2}\,{\rm eV}$ where axions could still
exist. They would then contribute most or all of the dark matter of
the universe. The ongoing direct search experiments for galactic
axions have reached a sensitivity where they are in a position to
confirm this bold hypothesis or to refute it.

\section*{Acknowledgments}

This work was partially supported by grant No.~SFB 375 of the Deutsche
Forschungsgemeinschaft.

%%%%%%%%%%%%%%%%%%%%%%%%%%%%%%%%%%%%%%%%%%%%%%%%%%%%%%%%%%%%%%%%%%%%%%
%% References %%%%%%%%%%%%%%%%%%%%%%%%%%%%%%%%%%%%%%%%%%%%%%%%%%%%%%%%
%%%%%%%%%%%%%%%%%%%%%%%%%%%%%%%%%%%%%%%%%%%%%%%%%%%%%%%%%%%%%%%%%%%%%%

\end{document}